**Title: Giant Thermal Conductivity Enhancement in Multilayer MoS$_2$ under Highly Compressive Strain**

*Xianghai Meng[†], Tribhuwan Pandey[†], Suyu Fu, Jing Yang, Jihoon Jeong, Ke Chen, Akash Singh, Feng He, Xiaochuan Xu, Abhishek K. Singh, Jung-Fu Lin\*, Yaguo Wang\**

[†]These authors contributed equally.

X. Meng, J. Jeong, Dr. K. Chen, F. He, Prof. Y. Wang

Department of Mechanical Engineering

The University of Texas at Austin

204 E. Dean Keeton St, Austin, TX 78712, USA.

E-mail:yaguo.wang@austin.utexas.edu

Dr. T. Pandey, A. Singh, Prof. A. K. Singh

Materials Research Centre

Indian Institute of Science

Bangalore 560012, India.

S. Fu, J. Yang, Prof. J. -F. Lin

Jackson School of Geosciences

The University of Texas at Austin

2305 Speedway Stop C1160, Austin, TX 78712, USA.

E-mail: afu@jsg.utexas.edu

F. He, Prof. Y. Wang

Texas Materials Institute

The University of Texas at Austin

204 E. Dean Keeton St, Austin, TX 78712, USA.

Dr. X. Xu

Omega Optics, Inc.

8500 Shoal Creek Blvd., Bldg. 4, Suite 200, Austin, Texas 78757, USA.

Prof. J. -F. Lin

Center for High Pressure Science and Technology Advanced Research (HPSTAR)

Shanghai 201900, China.

Keywords: TMD, thermal conductivity, compressive strain, phonon

**Abstract**

Multilayer $MoS_2$ possesses highly anisotropic thermal conductivities along in-plane and cross-plane directions that could hamper heat dissipation in electronics. With about 9% cross-plane compressive strain created by hydrostatic pressure in a diamond anvil cell, we observed about 12 times increase in the cross-plane thermal conductivity of multilayer $MoS_2$. Our experimental and theoretical studies reveal that this drastic change arises from the greatly strengthened interlayer interaction and heavily modified phonon dispersions along cross-plane direction, with negligible contribution from electronic thermal conductivity, despite its enhancement of 4 orders of magnitude. The anisotropic thermal conductivity in the multilayer $MoS_2$ at ambient environment becomes almost isotropic under highly compressive strain, effectively transitioning from 2D to 3D heat dissipation. This strain tuning approach also makes possible parallel tuning of structural, thermal and electrical properties, and can be extended to the whole family of 2D Van der Waals solids, down to two layer systems.

**Main Text**

Molybdenum disulfide ($MoS_2$), one member in transition metal dichalcogenides (TMDs) family, has received tremendous attention for its unique electronic, optical, and mechanical properties. While most recent research focuses on monolayer $MoS_2$, which has an appreciable direct band gap of about 1.9 eV,[1] multilayer $MoS_2$ attracts significant interests for various applications in nanoelectronic,[2] optoelectronics,[3] and flexible devices,[4,5] due to its high electron density of states,[6] and wider spectrum response.[7] $MoS_2$ possesses a two-dimensional (2D) layered structure where individual layers stack to form a multilayer crystal. Intra-layer atoms in $MoS_2$ are connected by covalent bonding, while inter-layer atoms by weak van der Waals (vdW) force. As a result, multilayer $MoS_2$ possesses highly anisotropic thermal conductivities along in-plane and cross-plane directions. Specifically, the reported in-plane thermal conductivity ranges from 35 to 85 W $m^{-1}K^{-1}$,[8-10] more than 10 times higher than the cross-plane thermal conductivity (less than 2.5 W $m^{-1}K^{-1}$).[8,11,12] For electronic application, high thermal conductivity along cross-plane direction, which also has large surface area, is preferred to remove heat efficiently. For most vdW 2D materials, the anisotropic thermal conductivity could seriously limit their ability of heat dissipation in 3D. Strain has been shown effective to tune physical properties in a wide range of materials,[13-15,16] however, experimental investigation of strain-tuned thermal conductivity in TMDs has not been reported. With traditional approaches, only a strain of less than 3% can be reached.[17,18] Hydrostatic pressure created in a diamond anvil cell (DAC) provides a convenient and effective means to realize compressive strain by as much as 30%.[19,20] Previous studies have shown that, with about 9% cross-plane compressive strain, multilayer $MoS_2$ exhibits a semiconductor to metal (S-M) transition,[21,22] with electrical conductivity enhanced by about 4 orders of magnitude.[19,23] The reduced inter-layer distance should also have a profound impact on phonon transport properties, which can greatly affect the cross-plane thermal conductivity in MoS$_2$. Exploring the tunability of thermal conductivity in multilayer TMDs with compressive strain, will benefit all TMD-based electronic devices.

In this work, we integrate time-domain thermoreflectance (TDTR) measurements with DAC to study strain-tuned cross-plane thermal conductivity in multilayer MoS$_2$ up to 25 GPa (11% strain along cross-plane direction). We have observed about a 12-time enhancement of cross-plane thermal conductivity. First-principle calculation and electrical conductivity measurements reveal that this drastic change mainly arises from the substantially strengthened inter-layer force and heavily modified phonon dispersions along the cross-plane direction. Measured with coherent phonon spectroscopy (CPS), the group velocities of coherent longitudinal acoustic phonons (LAP) are found to be doubled at 20 GPa due to phonon hardening, while their lifetimes decrease due to phonon unbundling. First-principle calculations also reveal that the anisotropic thermal conductivity in multilayer MoS$_2$ at ambient pressure becomes almost isotropic under highly compressive strain, showing a 2D to 3D transition. Even though we have applied extreme strain condition to reveal the limit of thermal conductivity tunability, what's more important is the sensitivity of thermal conductivity with strain in 2D materials. For example, under only 3% compressive strain, which is achievable mechanically, a 5~6 times enhancement has been reached. These findings can be extended to any 2D material systems bonded by VdW forces, down to two-layer systems.

As shown in **Figure 1a**, a diamond anvil cell (DAC) is implemented into a femtosecond pump-probe spectrometer for both time-domain thermoreflectance (TDTR) and coherent phonon spectroscopy (CPS) measurements. The force exerted onto two opposing diamonds is transformed into uniform hydrostatic pressure through the Neon medium, with the pressure monitored by a ruby calibrant placed close to the sample. (See Methods and Supporting Information for more details)

A multilayer MoS$_2$ sample with a thickness of about 1.5 μm is used for TDTR measurements, which is prepared by mechanical exfoliation and sonication (see Methods). Raman spectroscopy (Figure S1) shows that our sample is in 2-H phase. As shown in Figure 1b, a 50 nm Cu thin film is deposited on top of MoS$_2$ sample to increase laser energy absorption and reflection, and the whole sample is loaded into DAC. Thermal reflectance spectra with a time delay up to 1.5 ns are collected at each pressure, two of which are presented in Figure 1c. The reflectivity over the full time range is plotted as inset of Figure 1c. The fast increasing and decaying components correspond to electron excitation and relaxation in Cu layer, and the slowly decaying component originates from electron-phonon interaction and thermal conduction. To exclude the complex electron-phonon coupling and make sure only thermal conduction is under investigation, only data after 200 ps time delay is used to extract thermal conductivity. Since both our sample thickness and probe laser spot size are much less than the spot size of heating pump pulse (about 10 μm, 1/e$^2$ diameter), a one-dimensional (1D) thermal conduction model (Equation 1) is used to calculate the time-resolved temperature profile, which is then compared with experimental data to extract thermal conductivity:

$$\rho c_p \frac{\partial T}{\partial t} = \kappa \frac{\partial^2 T}{\partial z^2} + S \qquad (1)$$

Where $\rho$ is density, $c_p$ is specific heat, $T$ is temperature, $\kappa$ is thermal conductivity and $S$ is source term by pump laser heating. With a Gaussian laser pulse, source term takes the form: $\dot{S} = I_0(1-R)\alpha \, exp\left(-\frac{t^2}{\tau^2}\right) exp(-\alpha z)$, where $I_0$ is peak laser intensity, $R$ is reflectivity of probe pulse, $\alpha$ is absorption coefficient of pump pulse (6.73×10$^5$ cm$^{-1}$ for Cu at 400 nm), and $\tau$ is laser pulse width (Full Width at Half Maximum, 370 fs at the sample position). With 0.12nJ pump pulse energy, the temperature rise in Cu layer is estimated to be around 11K.

With the sample configuration shown in Figure 1b, four materials have to be considered in thermal conduction model: Neon, the pressure medium; Cu film; MoS$_2$ and the diamond anvils. At room pressure, the thermal conductivity of Ne medium is only about 0.05 W m$^{-1}$K$^{-1}$,[24] much smaller than that of Cu and MoS$_2$, heat conduction into Ne can be neglected. At elevated pressure (>10GPa), Ne solidifies and its thermal conductivity will increase substantially. However, Due to the large interface thermal resistance between Cu and solid Ne (see Supporting Information Figure S5), the heat flow into solid Ne is still negligible within our time range even when its thermal conductivity reaches 50 W m$^{-1}$K$^{-1}$. The maximum time delay, $t_{max}$, used in TDTR measurements is about 1.5 ns. Within this time range, the thermal penetration depth $D = \sqrt{\frac{k}{\rho c_p} \cdot t_{max}}$, is estimated to be about 210 nm in Cu, with $\kappa_{Cu}$=100 W m$^{-1}$K$^{-1}$, $\rho_{Cu}$=8,960 kg m$^{-3}$, and $c_{p\_Cu}$=0.39 J g$^{-1}$K$^{-1}$. Because $D_{Cu}$ is much larger than Cu layer thickness (50 nm), heat can penetrate through Cu layer and well into the MoS$_2$ sample, hence measurement in this time range should be very sensitive to the thermal properties of MoS$_2$. For MoS$_2$, even with $\kappa_{\perp MoS2}$=10 W m$^{-1}$K$^{-1}$, about 5 times larger than its ambient value, $D_{MoS2}$ is estimated to be about 90 nm, with $\rho_{MoS2}$=5,060 kg m$^{-3}$ and $c_{p\_MoS2}$=0.37 J g$^{-1}$K$^{-1}$. Because $D_{MoS2}$ is much smaller than MoS$_2$ thickness (1.5 μm), thermal conduction mainly happens within MoS$_2$ layer and does not have sufficient time to reach the MoS$_2$/diamond interface. As a result, a two-layer 1D heat conduction model only considering Cu and MoS$_2$ is adopted for analysis of TDTR spectra to obtain thermal conductivity.

Pressure could strongly affect not only $\kappa_{MoS2}$, but also copper density $\rho_{Cu}$, MoS$_2$ density $\rho_{MoS2}$, copper thermal conductivity $\kappa_{Cu}$ and the thermal interfacial resistance $R_{int}$ between Cu and MoS$_2$, which have to be examined carefully. Sensitivity studies with simulations (See Figure S8 in Supporting Information) show that temperature profiles from 200 ps to 1.5 ns are mostly sensitive to the change of $\rho_{Cu}$, $\rho_{MoS2}$ and $\kappa_{MoS2}$, which should be considered in our thermal conduction model. Density change of copper with pressure has been recently reported[25]. Dependence of $\rho_{MoS2}$ with pressure can be conveniently derived from the equation of state of MoS$_2$ published in ref. 19. (See

Figure S9 in Supporting Information) The insensitive parameters $\kappa_{Cu}$ is set to be 100 W m$^{-1}$K$^{-1}$.[26] $R_{int}$ is determined to be $2\times10^{-9}$ K W$^{-1}$m$^{-2}$ from the best fitting result (See Figure S6&S7 in Supporting Information).

Plotted in Figure 1d is the pressure-dependent cross-plane thermal conductivity, $\kappa_{\perp,total}$ of MoS$_2$ extracted from TDTR measurements. The top axis shows the corresponding strain rate along cross-plane direction, which is derived from the equation of state of MoS$_2$.[19] $\kappa_{\perp,total}$ undergoes a dramatic change from about 1 W m$^{-1}$K$^{-1}$ at ambient pressure to about 12 W m$^{-1}$K$^{-1}$ at 15 GPa, and tends to saturate thereafter. This drastically enhanced $\kappa_{\perp,total}$ of MoS$_2$ with pressure could have two possible origins: a) enhanced electronic contribution to thermal conduction along with S-M transition; b) enhanced phonon contribution due to reduced inter-layer distance and modified phonon structure.

Figure 2a shows the pressure-dependent electron thermal conductivity converted from previously measured electrical conductivity[19] using Wiedemann-Franz Law: $\kappa_e/\sigma=LT$, where $\sigma$ is electrical conductivity, $L$ is Lorenz number taken as $2.44\times10^{-8}$ WΩ K$^{-2}$,[27] and $T$ is temperature (300 K in here). Electron thermal conductivity $\kappa_e$ increases almost by 4 orders as pressure increases from ambient to 25 GPa, which is due to the S-M transition that starts around 10 GPa and finishes around 20 GPa.[19] When the electronic band gap of multilayer MoS$_2$ closes at high pressure, electrons at valence band can easily move into conduction band and become free electrons, which can greatly enhance both electrical conductivity and electronic thermal conductivity. However, compared with the pressure-dependent $\kappa_{\perp,total}$, contribution from $\kappa_e$ is extremely small, even for the metallic state. This is strikingly different from most metals, where electronic thermal conductivity dominates heat conduction. For metallic MoS$_2$,[19,28] small $\kappa_e$ is attributed to the small electron density of states near the Fermi Level. Therefore, the giant enhancement observed in $\kappa_{\perp,total}$ should be associated with the pressure-modified phonon properties.

First-principle calculations have been carried out to investigate effects of compressive strain on inter-/intra- layer interatomic forces and phonon properties. Figure 2b presents the trace of interatomic force constants (IFC) for three different types of bonds in multilayer $MoS_2$ (see Supporting Information for details of IFCs calculation). Because of the reduced inter-layer distance, at 20 GPa the inter-layer bonds Mo-S and S1-S2 have been strengthened by 1000% and 200%, respectively; on the contrary, intra-layer S-S bonds are slightly weakened with pressure. This anisotropic evolution of IFCs with pressure has profound impacts on phonon structure. Figure 2c shows phonon dispersions along both cross-plane (Γ-A) and in-plane (Γ-M, Γ-K) directions, with group velocities indicated as progressive colors. For $MoS_2$, the main contributions to thermal conduction come from three acoustic branches (2 transverse acoustic modes (TA), 1 longitudinal acoustic mode (LA)) and three low-frequency optical branches (2 transverse optical modes (TO), 1 longitudinal optical mode (LO)). At ambient pressure, the in-plane phonons possess much larger group velocities and phonon frequencies than cross-plane phonons; therefore, the in-plane thermal conductivity is much higher than its cross-plane counterpart. Moreover, cross-plane optical and acoustic branches are bundled into a narrow frequency range. When pressure goes up, frequencies and group velocities of all phonon branches along Γ-A direction rise rapidly (phonon hardening effect) and their dispersions spread into a broader frequency region (unbundling effect). In-plane phonons only show slight increase in frequencies and group velocities, and their dispersions split above 5 GPa (unbundling effect). At higher pressures (13.5 GPa and 19.9 GPa), in-plane optical branches are squeezed into a narrower frequency region and their group velocities decrease (phonon softening effect). Significant phonon hardening along cross-plane direction and modest phonon softening of optical phonons along in-plane directions are direct results of substantially strengthened inter-layer bonds (Mo-S, S1-S2) and slightly weakened intra-layer bond (S-S), respectively. These modifications on phonon structure have direct impacts on thermal conductivities.

Plotted in Figure 3a is the cross-plane lattice thermal conductivity $\kappa_{\perp,lattice}$, with contributions from two TA branches, one LA branch and all optical branches. Even though contributions from all phonon branches are enhanced by pressure, LA phonons dominant over the TA and optical phonons, which deserves further investigation. Coherent phonon spectroscopy (CPS) has been used to measure the pressure-dependent group velocity and lifetime of coherent longitudinal acoustic phonons (LAP) along the cross-plane direction. For CPS experiment, we use a bare multilayer $MoS_2$ sample with thickness about 1μm and size about 100 μm. When pump pulses are absorbed at the $MoS_2$ surface, a wave packet of coherent acoustic phonons are generated and propagate into the sample. The traveling coherent phonons can modify the local dielectric constants and cause partial reflection of the probe pulse (Impulsive Brillouin Scattering), which will interfere constructively or destructively with the reflected probe pulse from the sample surface and lead to the oscillations observed in Figure 3b. Signals from other components, such as electronic excitation and relaxation, have been removed with a digital filter. The solid curves in Figure 3b are fittings with damped harmonic oscillators:

$$\frac{dR}{R} = A \cdot exp(-\frac{t}{\tau_{LAP}}) \cdot cos(2\pi f t + \varphi) \qquad (2)$$

where $A$ is phonon amplitude, $\tau_{LAP}$ is phonon lifetime, $f$ is phonon frequency and $\varphi$ is the initial phase of phonon oscillations. Phonon frequency can be converted to phonon group velocity using the relation: $v_g = \lambda f/2n$, where $\lambda$ is the probe wavelength (800 nm), and $n$ is refractive index (4.2 for multilayer $MoS_2$).[29,30]

Figure 3c and 3d display the pressure-dependent phonon frequencies, group velocities, and phonon lifetimes of detected LAP from CPS experiments and first-principle calculations. Both experimental and simulation results show that with increasing pressure, group velocity of LAP increases more than 100% but phonon lifetime drops by 2~3 times. As discussed earlier, the increase of LAP group velocity is mainly a result of enhanced inter-layer interaction. The decrease of phonon lifetime could be related to the intensified three-phonon (anharmonic) scattering due to phonon unbundling at high pressure. Theoretical studies have shown that phonon bundling can greatly reduce three-phonon scatterings, because momentum and energy conservations could not be satisfied simultaneously.[31,32] Phonon unbundling under high pressure revealed in Figure 2c should give the opposite effect. Under high pressure, cross-plane optical and acoustic phonon branches spread into a broader frequency regime, providing more possible scattering channels, which can greatly reduce phonon lifetime. Results from first-principle calculations further confirm that anharmonic scatterings have significant increase with pressure (See Figure S11 in Supporting Information). The green symbols in Figure 3d show that the lifetime of $A_{1g}$ optical phonons also decreases under high pressure (derived from our previous Raman measurements in ref. 19), following a similar trend with LAP. Based on discussions above, it can be concluded that the 12-time increase of $\kappa_{\perp,total}$ with pressure mainly comes from strengthened inter-layer force and enhanced group velocity of LAPs. The saturation above 15 GPa is associated with the combined effects from increasing group velocity and reduced phonon lifetimes.

It is well known that $MoS_2$ possesses highly anisotropic thermal conductivity at ambient pressure.[9,11,33,34] Our study here reveals that the phonon group velocity and thermal conductivity along the cross-plane direction experience drastic increase with pressure, while modest changes have been observed for in-plane phonon properties. Figure 4a plots the first-principle calculations of in-plane lattice thermal conductivity $\kappa_{//,lattice}$, with contributions from 2TA, 1LA and all optical branches. Several features observed here distinguish from cross-plane phonons: i) optical phonons' contribution is comparable to that of acoustic phonons, due to their high group velocities; ii) contribution from LA phonons is comparable to that of TAs; and iii) acoustic phonon lattice thermal conductivity increases monotonically with pressure until 15 GPa, then drops slightly. Optical phonon lattice thermal conductivity increases slightly for pressures < 5G Pa and then decreases monotonically at higher pressures, which come from phonon softening and unbundling effect observed in Figure 2c. The overall $\kappa_{//,lattice}$ is plotted in Figure 4b; for comparison, the overall $\kappa_{\perp,lattice}$ is also plotted. At relatively low pressures (< 5 GPa), both $\kappa_{//,lattice}$ and $\kappa_{\perp,lattice}$ increase, mainly due to phonon hardening. At higher pressures, $\kappa_{\perp,lattice}$ continues to increase but $\kappa_{//,lattice}$ starts to drop, which makes them comparable at around 20 GPa. The anisotropy can be illustrated by plotting the thermal conductivity values along arbitrary crystallographic directions, as shown in Figure 4c. The disk-like shape at ambient pressure suggests that in-plane lattice thermal conductivity ($\kappa_{xx}$, $\kappa_{yy}$) is much greater than cross-plane counterpart ($\kappa_{zz}$), confirming the anisotropic thermal conduction from the 2D nature of multilayer MoS$_2$. At 20 GPa, the in-plane lattice constant $a_0$ has about 4.6% reduction compared to its ambient value, and the cross-plane lattice constant $c_0$ has about 10% reduction. The initial disk-like shape becomes almost spherical, indicating isotropic thermal conduction. These results suggest that pressure-induced change of the inter-layer force can significantly reduce and may eventually totally eliminate the anisotropy of thermal conductivity in multilayer MoS$_2$, effectively transitioning from 2D to 3D.

The most striking finding of this study centers around the observation of an about 12-time jump of $\kappa_{\perp,total}$ in multilayer MoS$_2$ under about 9% compressive strain, which drastically transitions the thermal conductivity from 2D to 3D. This transition in thermal conductivity has a different physical origin than that in the S-M transition reported previously. The giant increase of $\kappa_{\perp,total}$ is dominated by heavily modified phonon properties, especially those of LAPs, not by electronic contribution. S-M transition starts around 10GPa, where $\kappa_{\perp,total}$ has already increased from 1 W m$^{-1}$K$^{-1}$ at ambient pressure to about 8~9 W m$^{-1}$K$^{-1}$. Also, the increasing trend of $\kappa_{\perp,total}$ saturates above 15 GPa, where electrical conductivity has the most rapid change. Multilayer MoS$_2$ maintains its semiconducting nature before and during the early stage of S-M transition (< 15 GPa), even though the electrical conductivity has already been enhanced by 2~3 orders. Therefore, it is possible to tune electrical and thermal properties simultaneously with pressure to achieve both high thermal conductivity and high electrical conductivity, while still maintaining the semiconducting character of MoS$_2$. High $\kappa_{\perp,total}$ will ensure that heat generated in electronic devices can be dissipated effectively and timely, which will greatly improve device performance and stability.

The 2D to 3D transition in thermal conductivity observed in this study occurs at relatively high pressure above 20GPa with a strain > 10%, because, under hydrostatic compression, the multilayer MoS$_2$ sample is compressed along both in-plane and cross-plane directions. Even though $\kappa_\perp$ increases monotonically until the pressure reaches 15 GPa, $\kappa_{//}$ only increases with pressures below 5 GPa and then decreases at higher pressures. If applied with a cross-plane uniaxial force, where the MoS$_2$ sample only experiences cross-plane compressive strain and could expand freely along in-plane direction, its in-plane thermal conductivity would drop faster, according to the predicted negative correlation between in-plane bonding strength and cross-plane thermal conductivity[16]. Thus, the 2D to 3D transition in thermal conductivity can be expected to occur at much lower compressive strain rate. This highly tunable $\kappa_\perp$ and controllable 2D to 3D transition in the thermal conductivity also opens opportunities to fabricate a thermal modulator based on strain-tuned TMDs,[35] which can control intensity and direction of heat flux in a similar way to electrical current and light. Finally, it is conceivable that this transition should occur in most other 2D materials with vdW interlayer force under strain.[36,37] These findings can be applied in most 2D vdW solids to improve heat dissipation of 2D electronics under compressive strain.

**Experimental Section**

1. **Sample preparation**

Multilayer $MoS_2$ samples were prepared by exfoliation from a single crystal (purchased from 2D semiconductors) with non-residual semiconductor tape (from UltraTape). After exfoliation, we employed the sonication method to remove the $MoS_2$ flakes from the tape, and then transferred a flake onto a diamond culet with a micro-edge tip. After sonication, $MoS_2$ flakes had sizes usually ranging from 10 μm to 100 μm. Because of the gasket hole size (300 μm and 200 μm, respectively, in two measurements) and its shrinkage under high pressure, only flakes with sizes below 70 μm were loaded into the diamond anvil cell (DAC) for measurements. Each DAC device consists of two opposing diamond culets and a Re gasket, enclosing a small chamber filled with a pressure transmitting medium and a ruby sphere as the pressure reference. The use of the Ne pressure medium in the sample chamber ensured that the $MoS_2$ sample was hydrostatically compressed under pressure. By measuring the fluorescence from ruby spheres, hydrostatic pressure can be characterized with an uncertainty of less than 1 GPa. Photographs of our DAC device and samples in the DAC can be found in Supporting Information Figure S2, ruby fluorescence at selected pressures is shown in Figure S3.

2. **Time-domain Thermoreflectance (TDTR) and Coherent Phonon Spectroscopy (CPS)**

Both TDTR and CPS are conducted with our home-built two-color femtosecond pump-probe spectrometer. Schematic layout of our pump-probe spectrometer can be found in Supporting Information Figure S2. Our Ti: Sapphire Amplifier system (Spitfire ACE, Spectra Physics) generates laser pulses with 35 fs pulse width (full width at half maximum), 5 kHz repetition rate,

1.2 mJ pulse energy, and 800 nm central wavelength. Firstly, a beam expander is used to reduce laser spot size from 15 mm to 5 mm (diameter), and then the laser beam is split into two parts: a strong pump beam and a much weaker probe beam. For each beam, a Glan prism and half wave plate are used together to control the beam intensity and polarization. In our experiments, both the pump and probe beams each have a polarization parallel to optical table surface. A second harmonic generation crystal (Beta Barium Borate, BBO) is used to convert pump beam from 800 nm to 400 nm, after which a band-pass filter is used to remove residual 800 nm component. A linear translation stage with a step motor is used to control the optical path of probe beam, through which time delays are created between pump and probe pulses with a resolution of 14 fs. Both pump and probe beams are focused onto the sample collinearly by a 20X objective lens (Mitutoyo) (working distance 30.5 mm) with a spot size of 10 μm and 1.5 μm, $1/e^2$ diameter, respectively. The probe beam reflected from sample surface, together with a reference probe beam, is collected by a balance detector. A mechanical chopper is placed in the path of the pump beam, and Pre-amplifier and Lock-in Amplifier are used to extract data and to improve signal-to-noise ratio.

For TDTR measurement, a 50 nm Cu transducer layer is deposited on $MoS_2$ sample surface with E-beam evaporation under $10^{-5}$ Torr pressure. For CPS measurement, $MoS_2$ samples with bare surfaces are used.

**3. First-Principal Calculation**

The phonon Boltzmann transport formalism combined with first principles density functional theory was used to calculate the lattice thermal conductivity of 2H-$MoS_2$ under pressure. The iterative solution of phonon Boltzmann transport equation requires knowledge of second and third order interatomic force constants, which are calculated using density functional theory. The second-order interatomic force constants were calculated using a super cell size of 5 × 5 × 3 (450 atoms). The quality of calculated second-order force constants were validated by comparing the calculated phonon dispersion to the experimental data. For third-order force constants, we calculate forces on atoms in large enough supercell of $3 \times 3 \times 2$ (108 atoms) with various atomic displacements. To calculate third order force constants we considered interaction up to third-nearest neighbors. The second- and third- order IFCs were calculated with the finite-displacement approach using Vienna ab initio simulation package (VASP).[38-40] In order to obtain accurate forces and phonon frequencies, a high energy cutoff of 600 eV and strict energy convergence criterion of $10^{-8}$ eV with a **k**-point grid of $5 \times 5 \times 3$ were used. The calculated second- and third-order force constants give the information about three-phonon scatterings rates, phonon frequency and group velocity. Then, with all these information, the lattice thermal conductivity was calculated by solving phonon Boltzmann Transport Equations iteratively. For this purpose, Brillouin zone integrations was carried out on a well converged **q**-mesh of $33 \times 33 \times 19$ with scale broadening parameter of 0.3. The procedure of solving the Boltzmann transport equation is implemented in the ShengBTE code.[41-43]

**Acknowledgements**

The authors are grateful for the supports from National Science Foundation (CARER, Grant No. CBET-1351881; NASCENT, Grant No. EEC-1160494), and Department of Energy (SBIR/STTR, Grant No. DE-SC0013178). AKS, TP and AS thank the Materials Research Centre and Supercomputer Education and Research Centre of Indian Institute of Science for providing computing facilities. AKS, TP and AS acknowledge support from DST Nanomission.

**Figures**

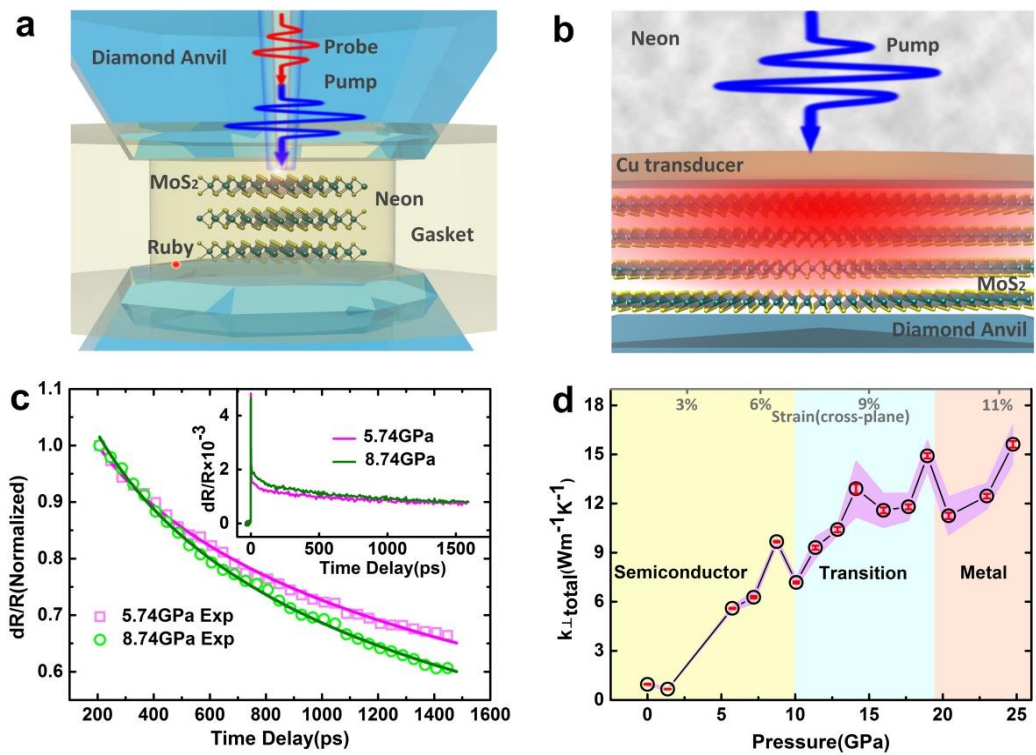

**Figure 1. Experimental setup and results of thermal conductivity measurement under high pressure. a)** Schematics of diamond anvil cell (DAC) with a multilayer $MoS_2$ sample installed, implemented into a femtosecond pump-probe spectrometer for both time-domain thermoreflectance (TDTR) and coherent phonon spectroscopy (CPS) measurements. **b)** Cross-section view of sample structure (Cu transducer layer deposited on multilayer $MoS_2$ sitting on a diamond surface) considered in thermal conduction model, heated with pump laser. **c)** Experimental data (open symbols) from TDTR measurements at selected pressures and the theoretical modeling (solid curves) with 1D thermal conduction model. Experimental data only after 200 ps time delay is used for thermal conductivity fitting. Inset shows experimental data over the whole time range. **d)** Extracted cross-plane thermal conductivity (both lattice and electronic) as a function of pressure. The shaded area marks for 90% confidence intervals from fitting. The three regions for semiconductor to metal transition given in ref. 19 are labeled.

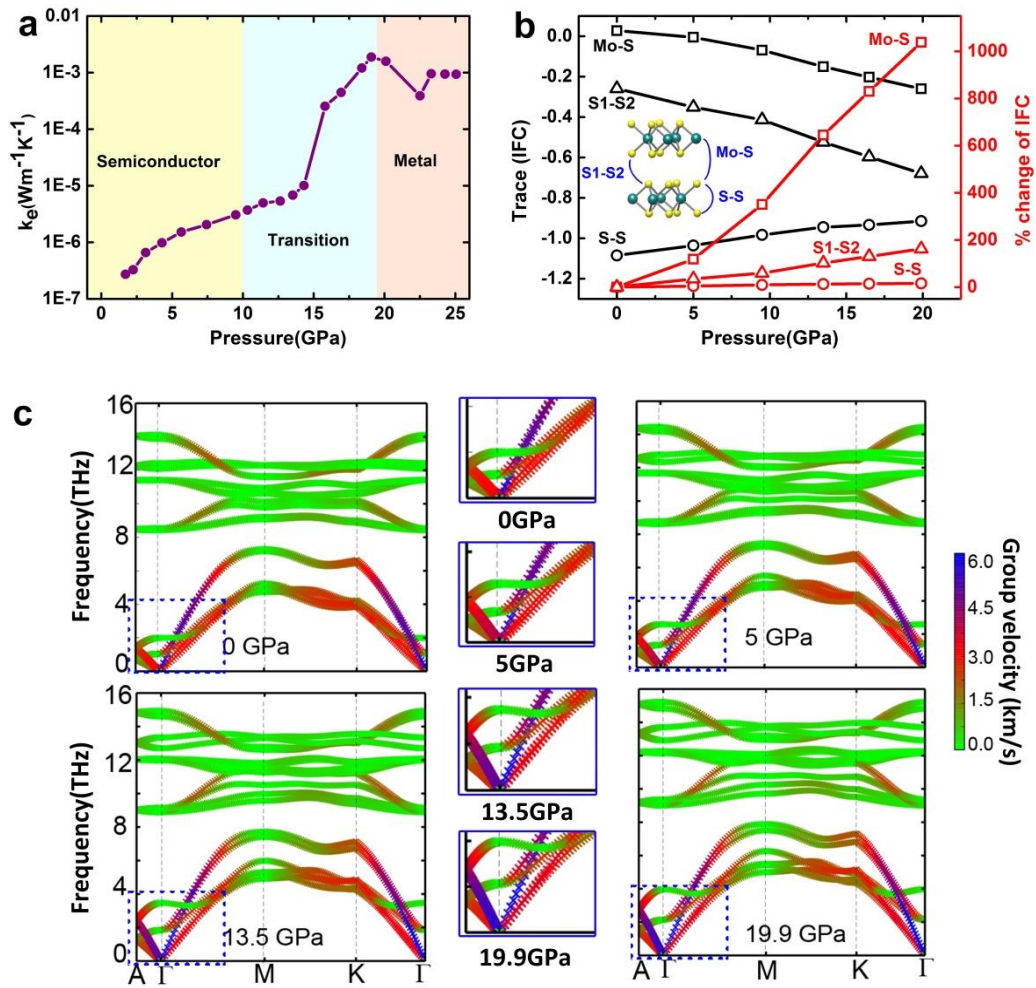

**Figure 2. Electronic thermal conductivity, interatomic force constant and phonon dispersion curves. a)** Electronic thermal conductivity of multilayer $MoS_2$ against pressure, converted with Wiedemann-Franz law from measured electronic conductivity. The three regions for semiconductor to metal transition given in ref. 19 are labeled. **b)** Pressure-dependent interatomic force constants of inter-layer Mo-S bond, inter-layer S-S (S1-S2) bond and intra-layer S-S bond, from first-principle calculations. **c)** First-principle calculations of pressure induced change of phonon dispersion curves and phonon group velocities in multilayer $MoS_2$, along both cross-plane (Γ-A) and in-plane (Γ-M, Γ-K) directions. Group velocities are presented with progressive colors. The middle panel are enlarged graphs for circled regions with dashed lines, showing details of phonon dispersions along Γ-A direction.

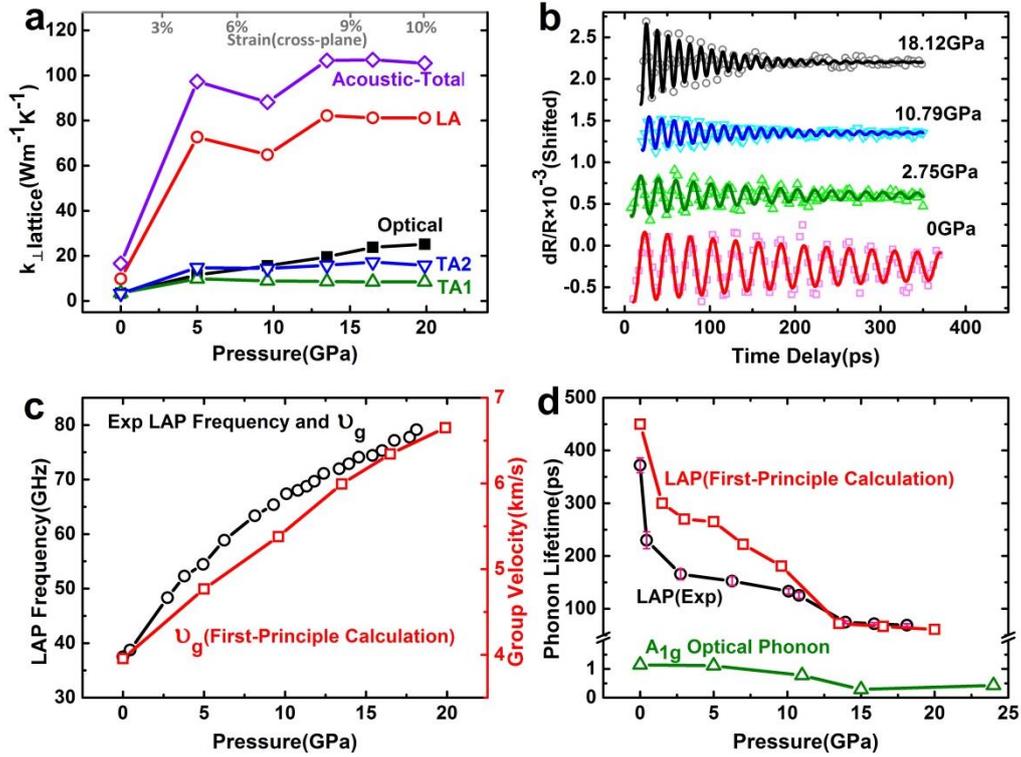

**Figure 3. First principles calculations of cross-plane thermal conductivity, CPS measurements of acoustic phonon frequency and lifetime. a)** Pressure-dependent cross-plane lattice thermal conductivity, obtained by first-principle calculations, with contributions from optical, longitudinal acoustic (LA), and two transverse acoustic (TA1, TA2) phonon branches. **b)** Pressure-dependent coherent oscillations of a longitudinal acoustic phonon (LAP) measured with CPS. Open symbols represent experimental data and solid curves represent fittings with damped harmonic oscillators. **c)** Pressure-dependent LAP frequencies (black circles) extracted from CPS measurements and LAP group velocities from first-principle calculations (red squares). **d)** Pressure-dependent LAP lifetimes (black circles) extracted from CPS measurements and from first-principle calculations (red squares), as well as lifetimes of $A_{1g}$ optical phonons (green triangles) extracted from Raman Measurements.[19]

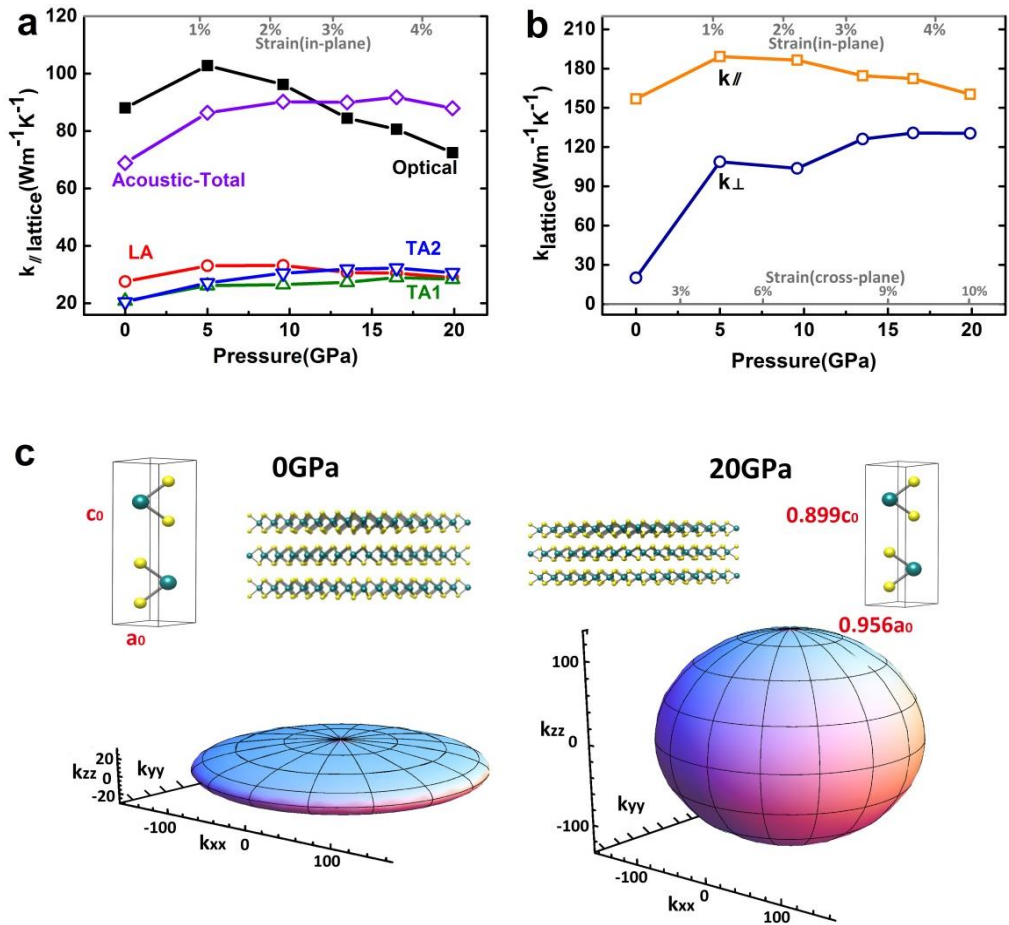

**Figure 4. 2D to 3D transition of thermal conductivity under high pressure. a)** Pressure-dependent in-plane lattice thermal conductivity calculated with first principles approach, with contributions from optical, longitudinal acoustic (LA) and transverse acoustic (TA1, TA2) phonons. **b)** Comparison of pressure-dependent cross-plane and in-plane lattice thermal conductivities. **c)** Schematic illustration of pressure-induced 2D to 3D thermal transition in multilayer $MoS_2$, with reductions of lattice constants along both in-plane and cross-plane directions.

Supporting Information

**Title**

**Giant Thermal Conductivity Enhancement in Multilayer MoS$_2$ under Highly Compressive Strain**

*Xianghai Meng[†], Tribhuwan Pandey[†], Suyu Fu, Jing Yang, Jihoon Jeong, Ke Chen, Akash Singh, Feng He, Xiaochuan Xu, Abhishek K. Singh, Jung-Fu Lin\*, Yaguo Wang\**

1. **Raman spectroscopy of multilayer MoS$_2$ sample**

As shown in Figure S1, Raman measurement of our bulk sample clearly shows E$_{2g}$ and A$_{1g}$ peaks. With increasing pressure, both peaks shift to larger wavenumbers, and the amplitude of E$_{2g}$ peak decreases. Both the peak positions at low and high pressure agree well with values reported in literature,[1] which confirms that our sample is in 2-H phase.

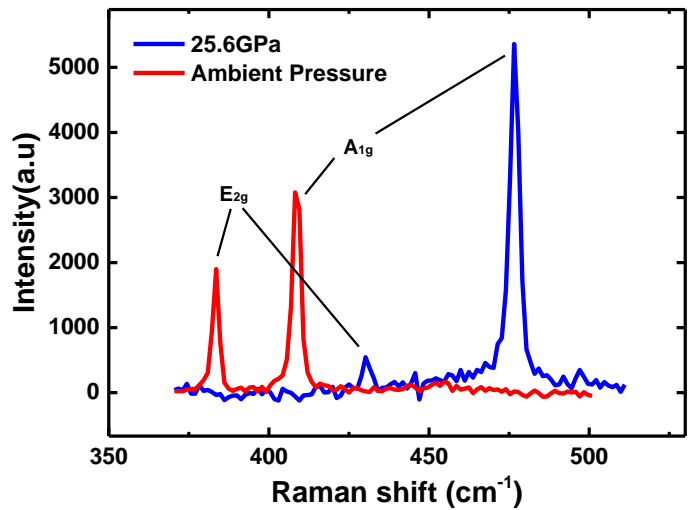

**Figure S1.** Raman shift of multilayer MoS$_2$ at ambient and high pressure

2. **Diamond Anvil Cell (DAC)**

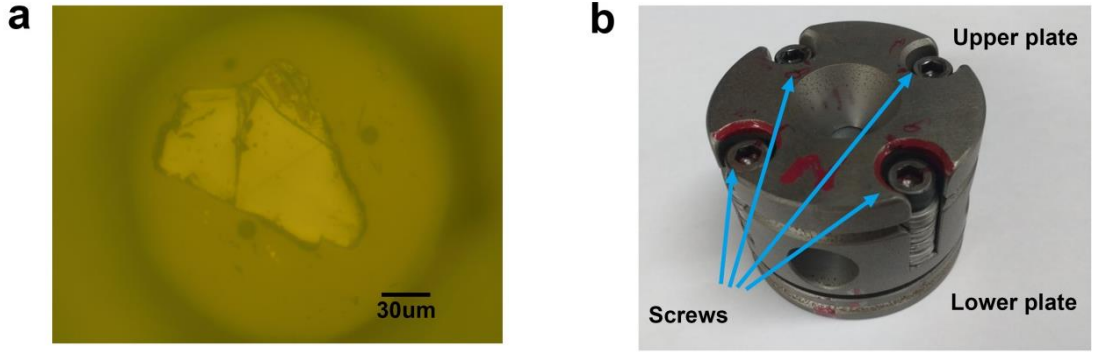

**Figure S2. a)** Optical image of multilayer MoS$_2$ in a diamond anvil cell. **b)** Picture of the cell device.

Figure S2a shows an optical image of a multilayer MoS$_2$ sample in DAC. The two small dark spheres on the lower left and upper right corners are rubies. The larger circle is the hole area that is drilled through the Re gasket. Figure S2b shows a picture of the DAC device with upper and lower plates connected by four screws. The DAC used for our experiments consists of two diamond culets with 500/400 μm surface area and a Re gasket, which was pre-indented to approximately 50 μm and consequently a hole was drilled in the Re gasket to form a small chamber filled with a multilayer MoS$_2$ sample,[1,2] the Ne pressure transmitting medium, and ruby spheres for pressure calibration. Diamonds sit on the diamond seats (center of the plates) that are glued onto upper and lower plates. Uniaxial pressure can be transformed into hydrostatic pressure by fastening the four screws evenly.

### 3. Pressure calibration with Ruby fluorescence

In Figure S3, we measure the fluorescence shift (R$_1$ peak) of Ruby loaded to DAC to monitor the pressure change,[4,5] which is a standard approach in high-pressure physics field. The equation below is use to determine the pressure based in the R$_1$ peak shift.

$$P = \frac{A}{B}\left\{\left[1 + \frac{\Delta\lambda}{\lambda_0}\right]^B - 1\right\} \quad \text{(Eq. S1)}$$

where A = 1920 GPa and B = 9.61 are coefficients, and $\lambda_0$ = 694.22 nm is the room-pressure value at 298 K.

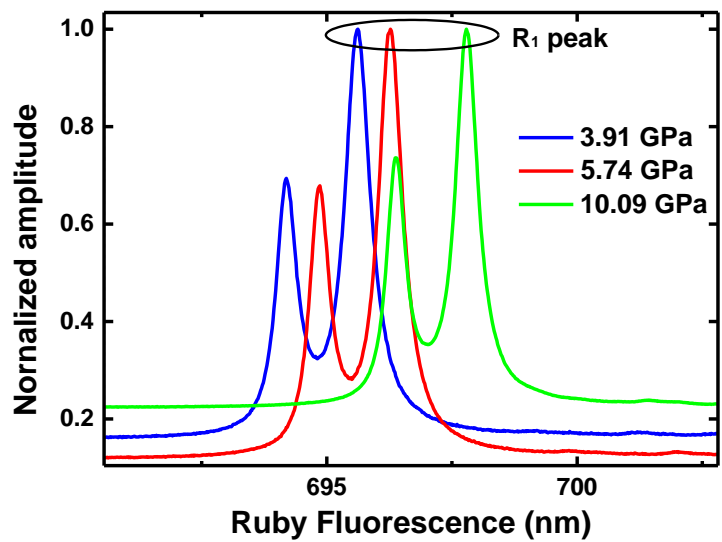

**Figure S3.** Ruby fluorescence $R_1$ peak shift with increasing pressure

## 4. Femtosecond pump-probe spectrometer

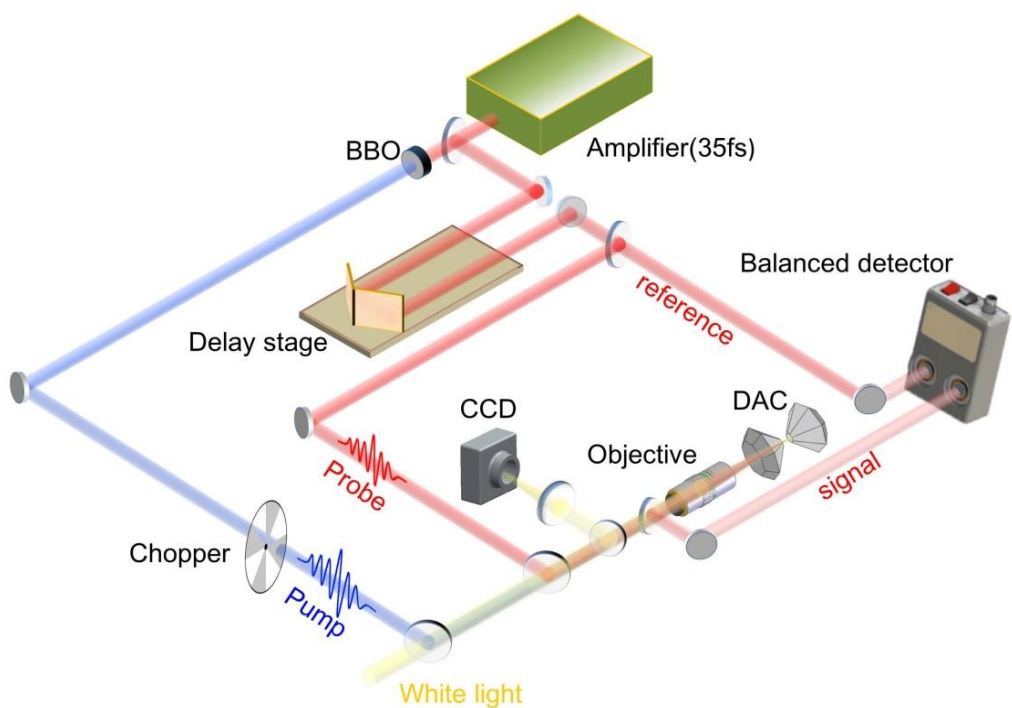

**Figure S4.** Schematics of pump-probe setup.

## 5. Effects from solid Ne medium and interface thermal resistance between Cu and solid Ne

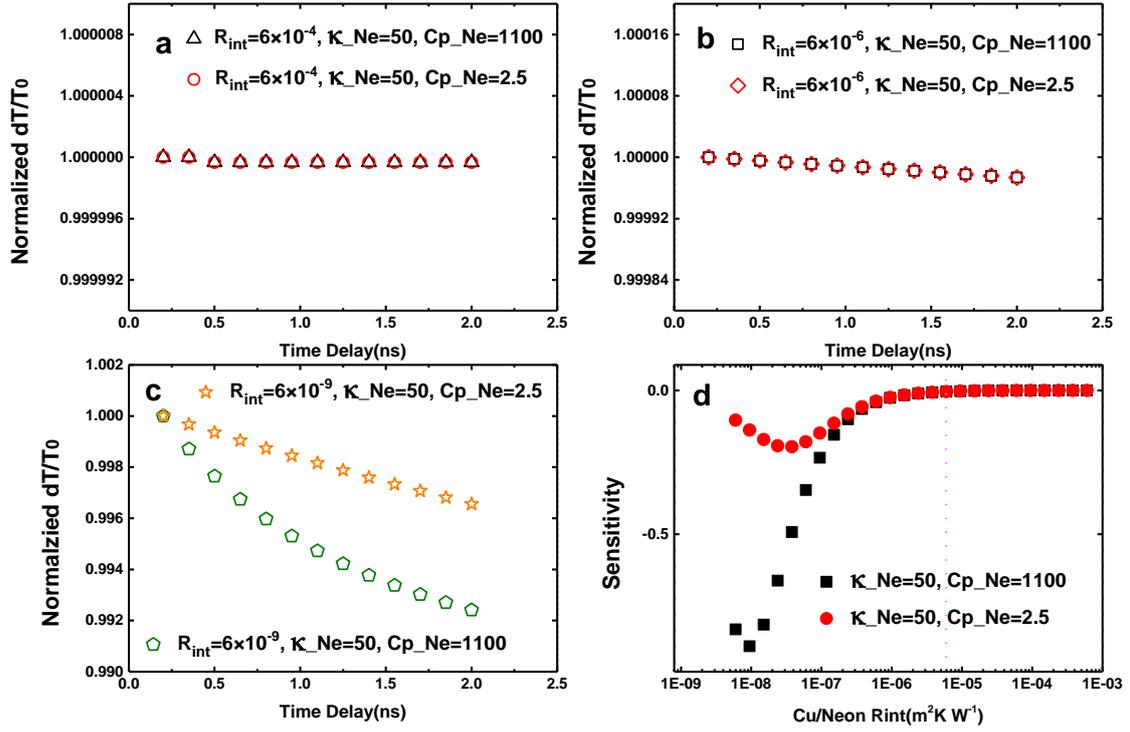

**Figure S5. a)-c)** Simulated curves of normalized temperature change $dT/T_0$ as a function of time delay under different Cu/Ne (solid) interface thermal resistances and specific heat of solid Ne ($R_{int}$ is in m$^2$KW$^{-1}$, $\kappa_{Ne}$ is in Wm$^{-1}$K$^{-1}$, $C_{p\_Ne}$ is in J Kg$^{-1}$K$^{-1}$). **d)** Sensitivity of temperature change with interface thermal resistance of Cu/Neon (solid) at 1ns time delay.

Thermal conductivity of solid neon at low temperature was reported to be about 20 Wm$^{-1}$K$^{-1}$ at around 6 K,[6] but no information is available in literature about solid Neon at high pressure. but no information is available in literature about solid neon at high pressure. Since both Argon and Neon are inert gases, we expect their thermal conductivities change with pressure follows a similar trend and we can scale the thermal conductivity of solid neon at high pressure from the well-documented Argon data. Argon thermal conductivity is estimated to increase almost linearly from 1.32 Wm$^{-1}$K$^{-1}$ at 1.66 GPa to 36.8 Wm$^{-1}$K$^{-1}$ at 24.5 GPa.[7] Since the solid neon thermal conductivity value 20 Wm$^{-1}$K$^{-1}$ is not obtained at high pressure but at low temperature, we estimate the maximum value at 25GPa to be about 50 Wm$^{-1}$K$^{-1}$, because neon solidifies at around 10 GPa. The lowest value of $\kappa_{Neon}$ is estimated to be about 1.8 Wm$^{-1}$K$^{-1}$ at 1.66 GPa, using the same ratio as $\kappa_{Argon}$.

The interface thermal resistance between solid Neon and Cu is reported to be about $6\times10^{-4}$ m$^2$KW$^{-1}$ at around 1.5 K.[8] The interface thermal resistances of Cu/D$_2$ (solid, isotope of Hydrogen) and Cu/H$_2$ (solid) are also shown to be about $12\times10^{-4}$ and $18\times10^{-4}$ m$^2$KW$^{-1}$, respectively, which are on the same order as Cu/Neon (solid). The interface resistance of Cu/Neon (solid) should be pressure dependent, but no information could be found in literature. Interface conductance of Al/SiO$_x$/SiC increases by about 10 times from 0 to 11GPa.[2] We expect the interface resistance of Cu/Neon (solid) will also decrease substantially with pressure. Literature has shown that thermal conductance between different interfaces can increase by 5 times from 70 K to 300 K.[9] The $R_{int}$ value of Cu/Neon we found in literature is at 1.5K, at room temperature, we expect the interface resistance decreases substantially. Considering both the pressure and temperature effects, and to be conservative, we assume the $R_{int}$ of Cu/Neon (solid) at 25GPa, 300K is about $6\times10^{-6}$ m$^2$KW$^{-1}$, 1% of its room pressure value at 1.5K.

Specific heat of solid neon also changes dramatically with pressure. Since $C_p$ value of solid neon at room temperature and high pressure is not available, we use the values measured at low temperature in solid neon. The specific heat of solid neon changes from about 2.5 J Kg$^{-1}$K$^{-1}$ at 2K to 1100 J Kg$^{-1}$K$^{-1}$ at around 20K.[10]

Summarized in Figure S5 a) ~ c) are the effects of $\kappa_{Ne}$, $C_{p\_Ne}$ and $R_{int}$ between Cu and solid Neon on temperature decay. We set $\kappa_{Ne}$ to be 50 Wm$^{-1}$K$^{-1}$, which is the maximum value at our pressure range. $C_{p\_Ne}$ is set to be 2.5 J Kg$^{-1}$K$^{-1}$ and 1100 J Kg$^{-1}$K$^{-1}$. Figure S5a, the simulated curves show that when the interface thermal resistance between solid neon and Cu is large ($6\times10^{-4}$ m$^2$KW$^{-1}$), effects from solid neon thermal conductivity and specific heat on the temperature change (dT/T$_0$ value) are negligible. As shown in Figure S5b, even the $R_{int}$ is reduced by 100 times, to be $6\times10^{-6}$ m$^2$KW$^{-1}$, effects from κ_Ne and C$_p$_Ne on the temperature decay are still negligible. Only when $R_{int}$ becomes extremely small, around $6\times10^{-9}$ m$^2$KW$^{-1}$, a clear temperature decrease could be observed, as shown in Figure S5c. Figure S5d shows a sensitivity study of temperature change as a function of R$_{int}$ between Cu and Neon, $d(Temperature)/d(ln(R_{int}))$, with solid neon thermal conductivity of 50 Wm$^{-1}$K$^{-1}$ and C$_p$ of 2.5 and 1100 J Kg$^{-1}$K$^{-1}$. It can be clearly seen that, when R$_{int}$ is larger than $6\times10^{-6}$ m$^2$KW$^{-1}$ (dashed line), the temperature change is not sensitive to the thermal conductivity and specific heat of solid neon.

In summary, these aforementioned analyses indicate that the heat dissipation through the Cu/solid neon interface and into the neon medium is negligible due to the large interface resistance, and the observed temperature change is dominated by heat dissipation through Cu and MoS$_2$ layers.

## 6. Effect from interface thermal resistance between Cu and MoS$_2$

Figure S6 shows the fittings of experimental data with two different interface thermal resistance between Cu and MoS$_2$, $2\times10^{-9}$ m$^2$KW$^{-1}$ and $5\times10^{-8}$ m$^2$KW$^{-1}$, which suggests a *R$_{int}$* on the order of $10^{-9}$ m$^2$KW$^{-1}$. With our measured values of MoS$_2$ thermal conductivity, R$_{int}$ of $2\times10^{-9}$ m$^2$KW$^{-1}$ gives a Kapitza length range about 1.94nm~25.8nm, much smaller than the thermal penetration depth in MoS$_2$ within 1.5ns, which is about 28.5~102.2nm. Hence the observed change of experimental data with pressure should not be dominated by Cu/MoS$_2$ interface thermal conductance. Figure S7 shows the sensitivity of Cu surface temperature change with MoS$_2$ thermal conductivity, $[\frac{d(T)}{d(ln(k_{MoS_2}))}]$, under different R$_{int}$ between Cu and MoS$_2$. It can be seen that the temperature change is very sensitive to MoS$_2$ thermal conductivity when R$_{int}$ is less than $1\times10^{-8}$ m$^2$KW$^{-1}$.

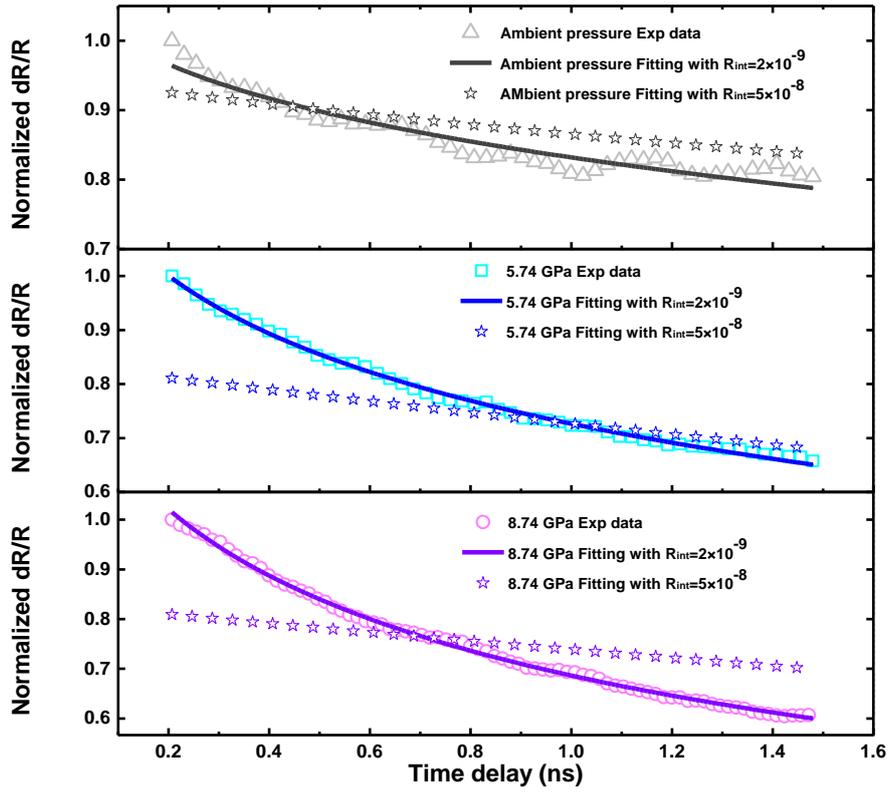

**Figure S6.** Experimental and fitted data at ambient pressure, 5.74GPa and 8.74GPa with Cu/MoS$_2$ interface thermal resistance of $2\times10^{-9}$ m$^2$KW$^{-1}$ and $5\times10^{-8}$ m$^2$KW$^{-1}$.

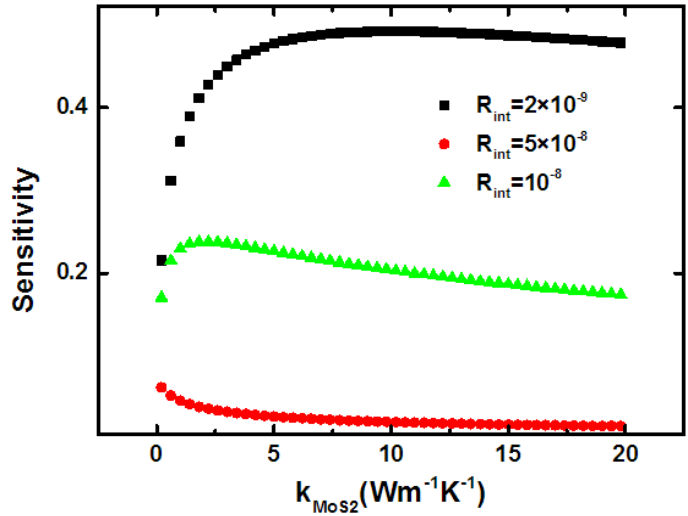

**Figure S7.** Sensitivity of temperature change with respect to thermal conductivity of MoS$_2$ under different Cu/MoS$_2$ interface thermal resistance ($R_{int}$ in m$^2$KW$^{-1}$) at 1ns time delay.

## 7. Sensitivity studies of fitting parameters in heat conduction model

In our 1D thermal conduction model to extract thermal conductivity from TDTR measurement, except $\kappa_{MoS2}$, several other parameters ($\kappa_{Cu}$, $R_{int}$, $\rho_{MoS2}$ and $\rho_{Cu}$) can change under high pressure, and need to be examined carefully in order to obtain accurate values for cross-plane $\kappa_{MoS2}$. Figure S8 presents simulations of time-resolved reflectivity change at $\kappa_{MoS2}$=1 W m$^{-1}$K$^{-1}$ and 10 W m$^{-1}$K$^{-1}$, with adjustments of $\kappa_{Cu}$, $R_{int}$, $\rho_{MoS2}$ and $\rho_{Cu}$ from their ambient values.

The first parameter we examine is thermal conductivity change of the Cu thermal transducer. Bridgman has shown that the thermal conductivity of Cu decreases by 10% from ambient pressure to 12 GPa,[11] then saturates onwards. Figure S8a shows that, with ±10% adjustment of $\kappa_{Cu}$ from its nominal value, 100 W m$^{-1}$K$^{-1}$, changes of dR/R curves are negligible. As a result, $\kappa_{Cu}$=100 W m$^{-1}$K$^{-1}$ is used for heat conduction model at all pressures. The second parameter we examine is the interfacial thermal resistance across Cu/MoS$_2$ interface. Hsieh et al. has shown that for a strong interface, such as metal/semiconductor, interface conductance change with pressure is minimal.[2] Figure S8b shows that, with ±10% adjustment of $R_{int}$ from ambient value, $2\times10^{-9}$KW$^{-1}$m$^{-2}$, changes of dR/R curves are negligible. As a result, $R_{int}=2\times10^{-9}$ K W$^{-1}$m$^{-2}$ is used for heat conduction model at all pressures.[12]

Figure S8c and S8d present sensitivity studies of $\rho_{Cu}$ and $\rho_{MoS2}$. With ±10% adjustment from their ambient values, changes of dR/R curves are substantial. As a result, changes of $\rho_{Cu}$ and $\rho_{MoS2}$ with pressure have to be considered in our heat conduction model. Pressure induced density changes of multilayer MoS$_2$ and copper are presented in Figure S9. The density change for multilayer MoS$_2$ is derived from equation of state published in ref. 1. The density change for copper is adapted from ref. 13.

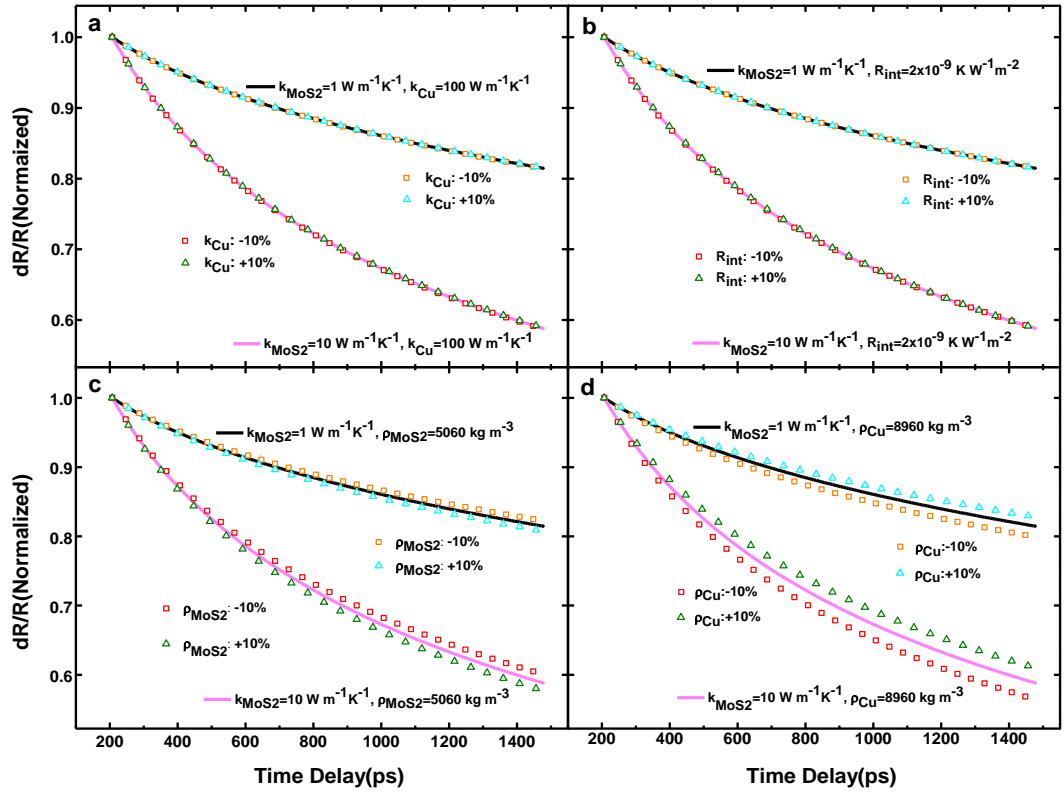

**Figure S8.** Sensitivity studies of fitting parameters used in thermal conduction model. **a)** Sensitivity of copper thermal conductivity change. Solid curves represent simulations of cases with $\kappa_{Cu}$=100 W m$^{-1}$K$^{-1}$ at $\kappa_{MoS2}$=1 W m$^{-1}$K$^{-1}$ (black) and 10 W m$^{-1}$K$^{-1}$ (purple). The open symbols show simulations where copper thermal conductivity is adjusted by ±10%. For all the cases presented: $R_{int}$=2×10$^{-9}$ K W$^{-1}$m$^{-2}$, $\rho_{MoS2}$=5060 kg m$^{-3}$, $\rho_{Cu}$=8960 kg m$^{-3}$. **b)** Sensitivity of interfacial thermal resistance $R_{int}$ change at Cu/MoS$_2$ interface. Solid curves represent simulations of cases with $R_{int}$=2×10$^{-9}$ K W$^{-1}$m$^{-2}$ at $\kappa_{MoS2}$=1 W m$^{-1}$K$^{-1}$ (black) and 10 W m$^{-1}$K$^{-1}$ (purple). The open symbols show simulations when interfacial thermal resistance is adjusted by ±10%. For all the cases presented: $\kappa_{Cu}$=100 W m$^{-1}$K$^{-1}$, $\rho_{MoS2}$=5060 kg m$^{-3}$, $\rho_{Cu}$=8960 kg m$^{-3}$. **c)** Sensitivity of MoS$_2$ density change. Solid curves represent simulations of cases with $\rho_{MoS2}$=5060 kg m$^{-3}$ at $\kappa_{MoS2}$=1 W m$^{-1}$K$^{-1}$ (black) and 10 W m$^{-1}$K$^{-1}$ (purple). The open symbols show simulations when MoS$_2$ density is adjusted by ±10%. For all the cases presented: $\kappa_{Cu}$=100 W m$^{-1}$K$^{-1}$, $R_{int}$=2×10$^{-9}$ K W$^{-1}$m$^{-2}$, $\rho_{Cu}$=8960 kg m$^{-3}$. **d)** Sensitivity of Cu density change. Solid curves represent simulations of cases with $\rho_{Cu}$=8960 kg m$^{-3}$ at $\kappa_{MoS2}$=1 W m$^{-1}$K$^{-1}$ (black) and 10 W m$^{-1}$K$^{-1}$ (purple). The open symbols show simulations when Cu density is adjusted by ±10%. For all the cases presented: $\kappa_{Cu}$=100 W m$^{-1}$K$^{-1}$, $R_{int}$=2×10$^{-9}$ K W$^{-1}$m$^{-2}$, $\rho_{MoS2}$=5060 kg m$^{-3}$.

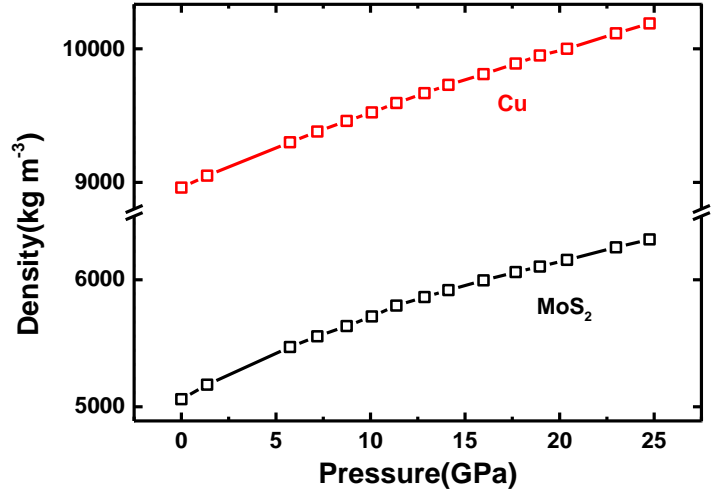

**Figure S9.** Pressure induced density change of multilayer $MoS_2$ (converted from equation of state in ref. 1) and Cu (adapted from ref. 13).

## 8. Extract properties of coherent longitidinal acoustic phonons (LAP)

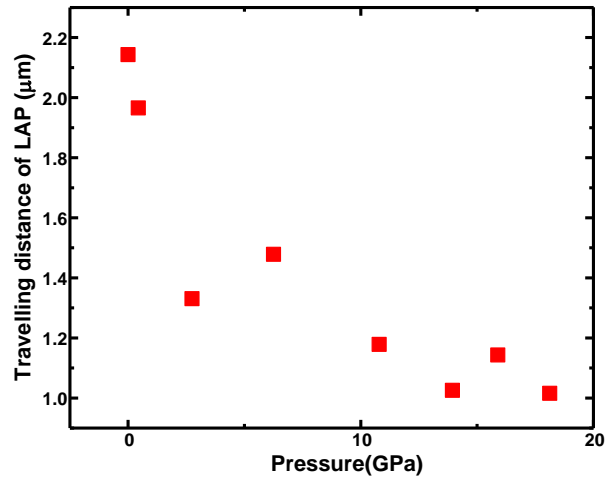

**Figure S10.** Travel distance of longitudinal acoustic phonons in multilayer $MoS_2$: $L=v_g \times \tau^*$, where $v_g$ is phonon group velocity and $\tau^*$ is the time when the last clear coherent phonon oscillation could be observed in experimental data (Figure 3b in main article).

Figure 3b in main article shows that phonon oscillations decay faster at high pressure. The observed phenomenon could also come from an artificial effect when the LAP travels out of the probing region of probe laser, or in order words, when the travel distance of the LAP is much longer than penetration depth of probe pulse in multilayer $MoS_2$. At ambient pressure, penetration depth of

800 nm light in multilayer $MoS_2$ is reported to be 5.88 μm or 11.76 μm,[14,15] which is much longer than our sample thickness (1.5 μm). At elevated pressures, penetration depth will reduce due to the change of refractive index. The authors are not aware of any published data about refractive index change of multilayer $MoS_2$ with pressure, so we can only estimate the change based on available information for other materials. Tsay et al.[16] calculated pressure dependence of refractive indices for common group IV and III-V compounds. From their results, at 20 GPa, the maximum refractive index change among all the materials simulated is about 23%. Dewaele et al[2]. measured refractive index change of $H_2$, He, $H_2O$ and Ne up to 35 GPa, and the maximum change they observed is about 30%. For multilayer $MoS_2$, if the refractive index is decreased by 30% at 20 GPa, the penetration depth will be about 4.1 μm (use 5.88 μm as ambient value), still much longer than our sample thickness.

Figure S10 plots the travel distance of LAPs estimated as: $L=v_g \times \tau^*$, where $v_g$ is phonon group velocity and $\tau^*$ is the time when the last clear coherent phonon oscillation could be observed in experimental data. It can be seen that at all pressures, travel distances of LAPs are much smaller than penetration depth of probe pulse. As a result, we can conclude with confidence that the faster decay of phonon oscillations observed in Figure 3b in main article does not come from the artificial effect, but do reflect reduction of phonon lifetime at high pressure.

## 9. Calculation of interatomic force constants (IFC)

To investigate the effect of hydrostatic pressure on strength of bonding, we calculated trace of IFC tensors with first-principles calculation:

$$\frac{\partial^2 E}{\partial R_I \partial R_J} = \begin{bmatrix} \frac{\partial^2 E}{\partial R_x \partial R_x} & \frac{\partial^2 E}{\partial R_x \partial R_y} & \frac{\partial^2 E}{\partial R_I \partial R_J} \\ \frac{\partial^2 E}{\partial R_y \partial R_x} & \frac{\partial^2 E}{\partial R_y \partial R_y} & \frac{\partial^2 E}{\partial R_y \partial R_z} \\ \frac{\partial^2 E}{\partial R_z \partial R_x} & \frac{\partial^2 E}{\partial R_z \partial R_y} & \frac{\partial^2 E}{\partial R_z \partial R_z} \end{bmatrix},$$  (Eq. s2)

where E and $R_i$ denote the energy and atomic position along $i^{th}$ direction ($i = x, y,$ and $z$), respectively. Bonding stiffness can be assessed with trace of IFC tensors, regardless of crystal structure or coordinate system:

$$\text{Trace of IFC} = \frac{\partial^2 E}{\partial R_x \partial R_x} + \frac{\partial^2 E}{\partial R_y \partial R_y} + \frac{\partial^2 E}{\partial R_z \partial R_z} \quad \text{(Eq. s3)}$$

In this study, we have calculated IFCs for S-S (intra-layer), Mo-S (inter-layer), and S1-S2 (inter-layer) bonds, which have been presented in Figure 2b in main article. At ambient condition, IFC of inter-layer Mo-S bond is positive due to the vdW interaction, which gives them an "anti-spring" behavior. With increasing hydrostatic pressure, IFCs of Mo-S bond becomes negative and increases by about 1000% at 20 GPa, indicating substantially strengthened interaction between interlayer atoms. Similarly, interlayer S1-S2 bond shows 200% increase of IFC at 20 GPa. In contrast, intra-layer S-S bond shows slight decrease from their ambient value, due to the strong covalent bonds among intra-layer atoms.

**10. Three-phonon scattering process as a function of pressure:**

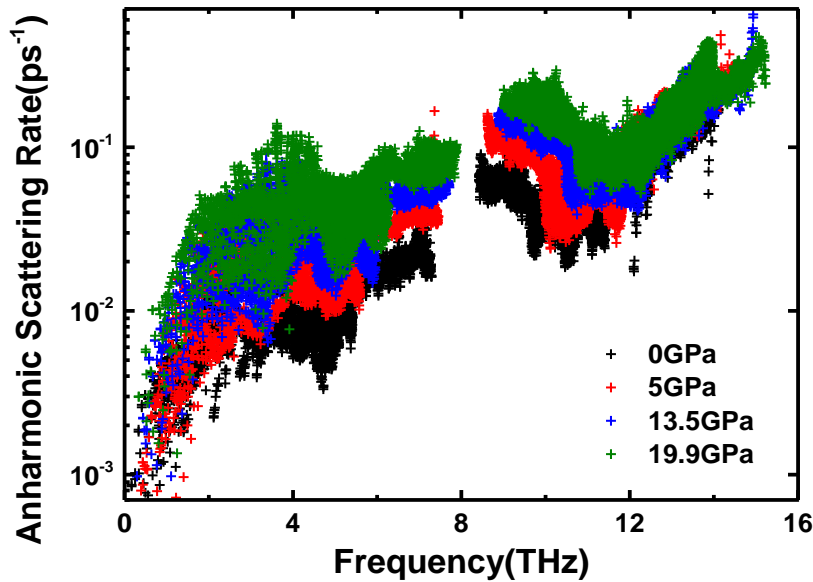

**Figure S11.** Frequency dependent anharmonic (three-phonon) scattering rates at various pressures

Figure S11 plots the anharmonic scattering rates for all phonon frequencies at various pressures.

Phonon scattering rate is the inverse of phonon lifetime. For almost all phonon frequencies (except for the highest frequency optical modes), anharmonic scattering is greatly enhanced at high pressure. Anharmonic (three-phonon) scattering requires both energy and momentum conservation. At ambient pressure, phonon dispersions are bundled together in a narrow frequency region (Figure 2c in main article), where energy conserved three-phonon scattering is suppressed. At elevated pressures, the phonon hardening effect pushes phonon dispersions to spread out into a broader frequency region (Figure 2c in main article), which increases energy conserved three-phonon scattering, as observed in Figure S11.

**11. Calculate thermal conductivity along arbitrary crystallographic directions**

Figure 4c in main article presents 3D illustrations of lattice thermal conductivity along arbitrary crystallographic directions in the first Brillouin zone. For example, the direction dependent lattice thermal conductivity can be represented by:

$$\kappa_{xyz} = \sum_{i=x,y,z}\sum_{j=x,y,z} \lambda_{xi}\lambda_{xj}\kappa_{ij} \qquad \text{(Eq. s4)}$$

Here $x$, $y$, and $z$ represents the Cartesian directions. $\lambda_{ij}$ represents directional cosines in spherical coordinates, i.e. $\lambda_{xx} = Sin\theta Cos\phi$, $\lambda_{xy} = Sin\theta Sin\phi$, and $\lambda_{xz} = Cos\theta$, with $\theta$ and $\phi$ being the inclination, and azimuth angles, respectively. For point group $D_{6h}$ (i.e MoS$_2$), all the cross components become zero and this equation can be rewritten as: $\kappa_{xyz} = \lambda_{xx}^2\kappa_{xx} + \lambda_{xy}^2\kappa_{yy} + \lambda_{xz}^2\kappa_{zz}$. By using the value of $\kappa_{xx}$, $\kappa_{yy}$ and $\kappa_{zz}$ in the above expression at a fixed temperature, lattice thermal conductivity can be obtained along arbitrary crystallographic direction. The three-dimensional lattice thermal conductivity contour plots are illustrated for 2H-MoS$_2$ for two different pressure values in Figure 4c. At 0 GPa, due to the hexagonal symmetry, thermal conductivity at the basal plane (x–y plane) is isotropic, and, because of the small value of the lattice thermal conductivity along the z direction, it takes an ellipsoidal shape. However, under pressure, due to the enhanced interaction between the layers, the lattice thermal conductivity becomes nearly isotropic as shown by the spherical distribution.